\documentclass[]{aa}  
\usepackage[varg]{txfonts}
\usepackage{natbib}
\usepackage{graphicx}

\begin{document}
\title{Search for magnetic fields in particle-accelerating colliding-wind
binaries\thanks{Based on archival observations obtained at the Telescope Bernard
Lyot (USR5026) operated by the Observatoire Midi-Pyr\'en\'ees, Universit\'e de
Toulouse (Paul Sabatier), Centre National de la Recherche Scientifique (CNRS) of
France, at the Canada-France-Hawaii Telescope (CFHT) operated by the National
Research Council of Canada, the Institut National des Sciences de l'Univers of
the CNRS of France, and the University of Hawaii, and at the European Southern
Observatory (ESO), Chile.}}
\titlerunning{Magnetic fields in PACWB}

\author{C. Neiner\inst{1}
\and J. Grunhut\inst{2}
\and B. Leroy\inst{1}
\and M. De Becker\inst{3}
\and G. Rauw\inst{3}
}

\offprints{C. Neiner}

\institute{LESIA, Observatoire de Paris, CNRS UMR 8109, UPMC, Universit\'e Paris
Diderot, 5 place Jules Janssen, 92190 Meudon, France; 
\email{coralie.neiner@obspm.fr}
\and European Southern Observatory (ESO), Karl-Schwarzschild-Str. 2, 85748,
Garching, Germany
\and Department of Astrophysics, Geophysics and Oceanography, University of
Li\`ege, 17 All\'ee du 6 Ao\^ut, B5c, 4000, Sart Tilman, Belgium
}

\date{Received ...; accepted ...}
 
\abstract
{Some colliding-wind massive binaries, called particle-accelerating
colliding-wind binaries (PACWB), exhibit synchrotron radio emission, which is
assumed to be generated by a stellar magnetic field. However, no measurement of
magnetic fields in these stars has ever been performed.}
{We aim at quantifying the possible stellar magnetic fields present in PACWB to
provide constraints for models.}
{We gathered 21 high-resolution spectropolarimetric observations of 9 PACWB
available in the ESPaDOnS, Narval and HarpsPol archives. We analysed these
observations with the Least Squares Deconvolution method. We separated the
binary spectral components when possible. }
{No magnetic signature is detected in any of the 9 PACWB stars and all
longitudinal field measurements are compatible with 0 G. We derived the upper
field strength of a possible field that could have remained hidden in the noise
of the data. While the data are not very constraining for some stars, for
several stars we could derive an upper limit of the polar field strength of the
order of 200 G.}
{We can therefore exclude the presence of strong or moderate stellar magnetic
fields in PACWB, typical of the ones present in magnetic massive stars. Weak magnetic fields could however be present in these objects.
These observational results provide the first quantitative constraints for
future models of PACWB.}


\maketitle

\section{Introduction}\label{intro}

Colliding-wind massive binaries (CWB) are binary systems composed of two stars
of O, early-B or WR type. Their main feature is a wind-wind interaction region
where the shocked gas is very hot (10$^7$ K). This wind interaction region is
likely to contribute to the thermal radio emission, in addition to the free-free
radiation due to thermal electrons in single star winds \citep{dougherty2003}.

In addition to this thermal emission, non-thermal radio emission was discovered
in some systems. It is related to the synchrotron radiation due to the presence
of relativistic electrons \citep{pittard2006}. These synchrotron emitters are
called particle-accelerating colliding-wind binaries (PACWB). In addition to the
synchrotron emission, these systems can be revealed by exceptionally large radio
fluxes, a spectral index significantly lower than the thermal value, and an
orbital modulation of the radio flux. \cite{debecker2013} recently provided the
most up-to-date catalog of such systems.

High angular resolution observations of some PACWB have allowed to disentangle
the thermal and non-thermal emissions \citep[e.g. OB2 \#5,][]{dzib2013} and
showed that the synchrotron emission is associated to the wind-wind interaction
region. This region is also a source of thermal X-rays, in addition to the
intrinsic X-ray emission produced in the stellar winds of the individual
components. The X-ray spectrum produced in the wind interaction region is
generally significantly harder than that of massive single stars, and the X-ray
emission is variable with the orbital phase
\citep[e.g.][]{debecker2011,cazorla2014}.

Finally, it was discovered more recently that PACWB may also emit $\gamma$ rays
through inverse Compton scattering by the relativistic electrons and neutral
pion decay. However, only one such example is known as of today
\citep[$\eta$\,Car,][]{farnier2011}.

These many characteristics make PACWB very interesting objects to study extreme
physical processes. However, it has become more and more clear over the last few
years that PACWB cover a very wide range of parameters (mass loss, wind
velocity, orbital period...) and the fundamental difference between the PACWB
and ``normal'' CWB is unknown \citep{debecker2013}.

The presence of synchrotron emission in PACWB immediately points towards the
presence of a magnetic field. Indeed, synchrotron emission results from the
modified movement of relativistic electrons in a magnetic field. Moreover, the
acceleration of particles in PACWB could be explained either by strong shocks in
the colliding winds \citep[e.g.][]{pittard2006} or by magnetic reconnection or
annihilation \citep[e.g.][]{jardine1996}. It has thus been speculated that the
fundamental difference between CWB and PACWB is the presence of a magnetic
field.

Over the last two decades magnetic fields have been detected in $\sim$7\% of
single massive stars \citep{wade2014}. While the fraction of PACWB among CWB is
not known and the catalog by \cite{debecker2013} certainly underestimates the
number of PACWB, $\sim$7\% could be a plausible proportion considering that only
43 possible PACWB have been identified as of today \citep{debecker2013} while
most massive stars are probably in binaries \citep{sana2012,sana2014}.
Therefore, the presence of a magnetic field might indeed be the difference
between PACWB and ``normal'' CWB.

The magnetic field in PACWB could be of stellar origin or it could also
possibly be generated in the colliding winds themselves. From synchrotron
observations, one can estimate the magnetic field strength in the wind-wind
interaction region to be of the order of a few mG \citep[see
e.g.][]{dougherty2003}. Extrapolating to the surface of the stars with typical
distances between the stagnation point and the photosphere, we obtain values of
the stellar magnetic field strength between one G and a few thousands G,
depending on the system and on the assumptions \citep[e.g.][]{parkin2014}.
Magnetic fields detected in single massive stars have a polar field strength
between hundred and several thousands G \citep[see e.g.][]{petit2013}, which
are compatible with the fields speculated in PACWB models.

Therefore, measuring magnetic fields in PACWB is an ideal way to test these
assumptions, constrain models of colliding winds, and understand the
difference between PACWB and ``normal'' CWB.

\section{Archival spectropolarimetric observations}\label{obs}

An updated census of 43 PACWB has been published recently \citep{debecker2013}.
It includes clear PACWB detected through their synchrotron emission as well as
candidates from indirect indicators (e.g. radio flux).

We have gathered all high-resolution spectropolarimetric data of these PACWB
available in archives, i.e. observed with Narval at T\'elescope Bernard Lyot
(TBL) in France, ESPaDOnS at the Canada-France-Hawaii telescope (CFHT) in
Hawaii, or HarpsPol at ESO in Chile. Circular polarisation data are available
for 9 of the 43 known PACWB. When several consecutive spectra were available for
the same night, we averaged them. The 9 stars and 21 (average) observations are
listed in Table~\ref{targets}.

\begin{table}
\caption{List of 21 archival spectropolarimetric observations of 9 PACWB,
including the instrument used for the observations, date of observations and
signal-to-noise ratio (SNR) in the Stokes I and V spectra.}
\label{targets}
\begin{tabular}{lllll}
\hline
\hline
Star & Instrument & Date & SNR I & SNR V\\
\hline
\object{HD\,36486}  &  Narval   & 23.10.2008 & 5386 & 21947  \\
	   &  Narval   & 24.10.2008 & 6021 &108758 \\
\object{HD\,37468}  &  ESPaDOnS & 17.10.2008 & 3149 & 56388  \\
\object{HD\,47839}  &  Narval   & 10.12.2006 & 4416 & 19648  \\
	   &  Narval   & 15.12.2006 & 4528 & 37218  \\
	   &  Narval   & 09.09.2007 & 4497 & 25269  \\
	   &  Narval   & 10.09.2007 & 4384 & 33820  \\
	   &  Narval   & 11.09.2007 & 4280 & 24183  \\
	   &  Narval   & 20.10.2007 & 4545 & 39090  \\
	   &  Narval   & 23.10.2007 & 4563 & 41668  \\
	   &  ESPaDOnS & 02.02.2012 & 4863 & 51425  \\
\object{HD\,93250}  &  HarpsPol & 17.02.2013 & 4169 &  9523  \\
\object{HD\,151804} &  HarpsPol & 26.05.2011 & 6191 & 22047  \\
\object{HD\,152408} &  ESPaDOnS & 05.07.2012 &  843 & 12909  \\
\object{HD\,164794} &  ESPaDOnS & 19.06.2005 & 3083 & 14933  \\
           &  ESPaDOnS & 20.06.2005 & 3298 & 15249  \\
           &  ESPaDOnS & 23.06.2005 & 3118 & 15657  \\
           &  HarpsPol & 25.05.2011 & 5050 & 14081  \\
	   &  ESPaDOnS & 14.06.2011 & 3346 & 29046  \\
\object{HD\,167971} &  ESPaDOnS & 30.06.2013 & 2671 & 22295  \\
\object{HD\,190918} &  ESPaDOnS & 25.07.2010 & 4096 & 13892  \\
\hline
\end{tabular}
\end{table}

For each star, we normalized the data to the intensity continuum level and
extracted Stokes $V$ and Null ($N$) polarisation spectra. $N$ spectra allow us
to check that the magnetic measurements (in the Stokes $V$ spectra) have not
been polluted by spurious signal, e.g. due to instrumental polarisation.

We then proceeded to use the Least Squares Deconvolution (LSD) technique
\citep{donati1997} to search for weak Zeeman signatures in the mean Stokes $V$
profile. The input LSD masks for each star were extracted from line lists
provided by VALD \citep{piskunov1995, kupka1999} according to the spectral type
of each target. These line lists orginally contain all lines with predicted line
depths greater than 1\%, assuming solar abundances. We proceeded to remove all
hydrogen lines, lines that were blended with H lines, and lines that are
strongly contaminated by telluric regions. We then automatically adjusted the
line depths of each remaining line to provide the best fit to the observed
Stokes $I$ spectra.

Using these final line masks, a mean wavelength of 5000 \AA\ and a mean Land\'e
factor of 1.2, we extracted LSD Stokes $I$ and $V$ profiles for each
spectropolarimetric measurements. We also extracted LSD $N$ polarisation
profiles to check for spurious signatures. All LSD $N$ profiles are flat,
showing that the LSD $V$ measurements do reflect the stellar magnetic field. The
LSD $I$ and Stokes $V$ profiles of the 9 stars are shown in Fig.~\ref{LSDprof}.
LSD $V$ profiles are also flat, showing no sign of a magnetic signature in any
of the 9 PACWB.

\section{LSD $I$ profile fitting}

To go further and evaluate the magnetic field in the studied PACWB, since
PACWB are binary stars, we first needed to separate the individual spectra of
each component in the LSD $I$ profiles.

For each spectrum we fit the mean LSD Stokes $I$ profile to determine the radial
velocity $V_{\rm rad}$, the projected rotational broadening ($v\sin i$) and any
contribution from non-rotational broadening that we consider to be
macroturbulent broadening ($V_{\rm mac}$).

\begin{figure*}[!ht]
\begin{center}
\resizebox{0.99\hsize}{!}{\includegraphics[clip]{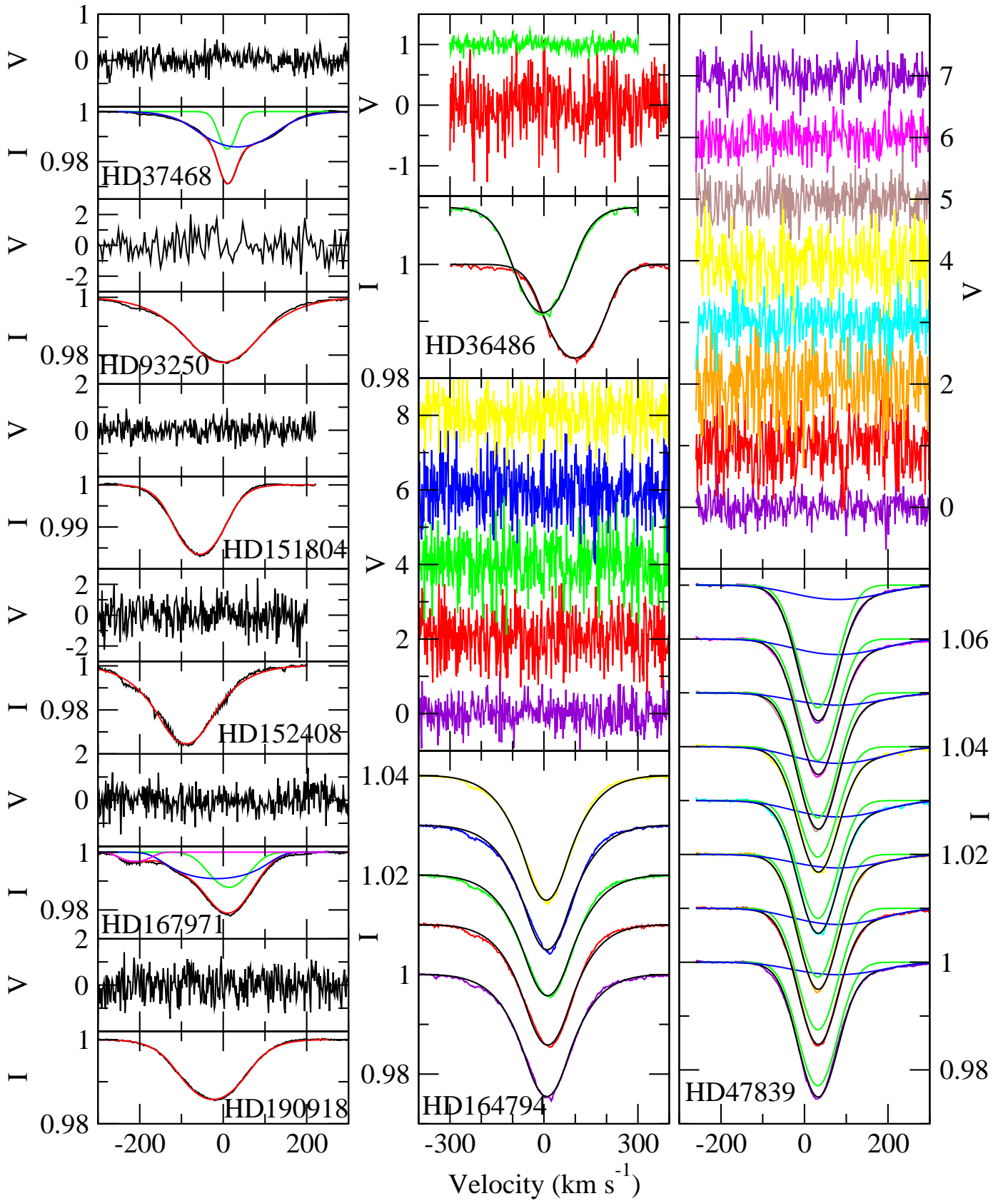}}
\caption[]{LSD $I$ (bottom panels) and Stokes $V$ (top panels, y-axis multiplied
by 10000) profiles of the 9 PACWB available in spectropolarimetric archives.
When only one spectrum is available (left panels), observations are indicated in
black and the fit in red. If several components are present the fit of the
primary/secondary/tertiary component is shown in blue/green/pink, while the
combined fit is in red. When several spectra are available for one star (middle
and right panels), observations are indicated with various colours, and the fits
are indicated in black with primary and secondary component fits indicated in
green and blue respectively. In these cases, profiles are artificially shifted
upwards to ease the reading.}
\label{LSDprof}
\end{center}
\end{figure*}

\begin{table}[!ht]
\caption{Parameters derived from the fit of LSD $I$ profiles of each star. When
several components are visible in the spectra, each component (primary,
secondary and possibly tertiary) is fitted. The last column indicates the upper dipolar field strength limit in G for each spectrum.}
\label{fit}
\begin{tabular}{lll@{\,\,}l@{\,\,}l@{\,\,}l@{\,\,}l}
\hline
\hline
Star & Date & Comp. & $V_{\rm rad}$ & $v\sin i$   & $V_{\rm mac}$ & $B_{\rm pol,max}$ \\
HD   &      &       & km~s$^{-1}$   & km~s$^{-1}$ & km~s$^{-1}$   & G \\
\hline
36486  &  23.10.08 &	  &  97 & 126 & 101 &  206 \\
       &  24.10.08 &	  &  -4 & 116 & 126 &  906 \\
37468  &  17.10.08 & prim &  33 & 115 & 124 &  258 \\
       &	   & sec  &  10 &  28 &  29 &  513 \\
47839  &  10.12.06 & prim &  32 &  52 &  90 &  867 \\
       &  	   & sec  &  79 & 140 & 154 & 8579 \\
       &  15.12.06 & prim &  32 &  52 &  90 &  472 \\
       &  	   & sec  &  79 & 140 & 154 & 4979 \\
       &  09.09.07 & prim &  32 &  52 &  90 &  687 \\
       &  	   & sec  &  79 & 140 & 154 & 5581 \\
       &  10.09.07 & prim &  32 &  52 &  90 &  543 \\
       &  	   & sec  &  79 & 140 & 154 & 4104 \\
       &  11.09.07 & prim &  32 &  52 &  90 &  789 \\
       &	   & sec  &  79 & 140 & 154 & 5658 \\
       &  20.10.07 & prim &  32 &  52 &  90 &  449 \\
       &	   & sec  &  79 & 140 & 154 & 3797 \\
       &  23.10.07 & prim &  32 &  52 &  90 &  428 \\
       &	   & sec  &  79 & 140 & 154 & 3837 \\
       &  02.02.12 & prim &  32 &  52 &  90 &  337 \\
       &	   & sec  &  79 & 140 & 154 & 3637 \\
93250  &  17.02.13 &	  &-1.5 &  92 & 189 & 4367 \\
151804 &  26.05.11 &	  & -58 &  79 &  83 &  850 \\
152408 &  05.07.12 &	  & -91 &  69 & 154 & 1363 \\
164794 &  19.06.05 &	  &  12 &  76 & 180 & 1600 \\
       &  20.06.05 &	  &10.5 &  75 & 191 & 1671 \\
       &  23.06.05 &	  & 9.5 &  71 & 191 & 1572 \\
       &  25.05.11 &	  & 8.9 &  71 & 186 & 1765 \\
       &  14.06.11 &	  & 6.7 &  69 & 189 &  865 \\
167971 &  30.06.13 & prim &  13 &  63 &  73 & 1092 \\
       &	   & sec  & -16 & 146 &  62 & 1160 \\
       &	   & ter  &-212 &  52 &  36 & - \\
190918 &  25.07.10 &	  & -25 & 102 & 111 & 1960 \\
\hline
\end{tabular}
\end{table}

Ideally, fits to the observed profiles should be computed with Fourier
techniques \citep[e.g.][]{gray2005,simondiaz2014} directly on the intensity
profiles (rather than the LSD profiles). However, this is very time consuming
and not necessary here since the exact value of the parameters are not important
for our purpose. We only need a good fit to the LSD profiles. Therefore, the
profiles are computed as the convolution of a rotationally-broadened profile and
a radial-tangential broadened profile following the parametrisation of
\cite{gray2005}, assuming equal contributions from the radial and tangential
(RT) component. While this form of macroturbulence is not commonly used in the
study of early-type stars (typically a Gaussian profile is used to characterise
$V_{\rm mac}$), \cite{simondiaz2014} showed that it provides a good
agreement with the Fourier techniques.

The code uses the {\sc mpfit} library \citep{more1978,markwardt2009} to find the
best fit solution. Using a radial-tangential profile for $V_{\rm mac}$
tends to maximise $v\sin i$. Therefore the values we obtain for $v\sin i$ can be
considered as upper limits.

For profiles that show obvious signs of spectroscopic companions (HD\,37468,
HD\,47839 and HD\,167971) we simultaneously fit multiple profiles, one for each
component, to determine the overall best solution for the given SB2 (or SB3)
profile. For HD\,47839, since the various spectra show no significant
variations, the averaged profile was fitted. The simultaneous fitting of
multiple profiles for one spectrum is a difficult task and the solution is often
degenerate. We therefore attempted to constrain each fit based on previous
studies published in the literature whenever possible. For the other stars
(SB1), only one component was fitted.

\begin{table}[!t]
\caption{Measured longitudinal field $B_l$ and Null polarization $N_l$, with
their error bars $\sigma$. When several components are visible in the spectra,
the values were estimated for each component (primary, secondary and possibly
tertiary).}
\label{bl}
\begin{tabular}{llllll}
\hline
\hline
Star   & Date      & Comp. & $B_l$ & $N_l$ & $\sigma$ \\
HD     &           &       & G     & G   & G \\
\hline
36486  &  23.10.08 &	  & 43   & 13   & 38 \\
       &  24.10.08 &	  & 18   &  1   &  7 \\
37468  &  17.10.08 & prim & -3   & 35   & 19 \\
       &	   & sec  & 16   & -18  & 9 \\
47839  &  10.12.06 & prim & -27  & 17   & 25 \\
       &  	   & sec  & -260 & -173 & 210 \\
       &  15.12.06 & prim & -6   & 14   & 13 \\
       &  	   & sec  & 5    & 256  & 185 \\
       &  09.09.07 & prim & 3    & -13  & 20 \\
       &  	   & sec  & -243 & -173 & 210 \\
       &  10.09.07 & prim & 8    & 7    & 15 \\
       &  	   & sec  & -41  & 278  & 154 \\
       &  11.09.07 & prim & 14   & -14  & 23 \\
       &	   & sec  & -141 & -95  & 211 \\
       &  20.10.07 & prim & -9   & 4    & 13 \\
       &	   & sec  & -98  & 146  & 140 \\
       &  23.10.07 & prim & -1   & 23   & 12 \\
       &	   & sec  & -27  & -154 & 142 \\
       &  02.02.12 & prim & -6   & -4   & 10 \\
       &	   & sec  & 17   & 22   & 132 \\
93250  &  17.02.13 &	  & -24  & 175  & 135 \\
151804 &  26.05.11 &	  & -13  & -34  & 33 \\
152408 &  05.07.12 &	  & 2    & 60   & 34 \\
164794 &  19.06.05 &	  & -3   & -18  & 52 \\
       &  20.06.05 &	  & -2   & 44   & 52 \\
       &  23.06.05 &	  & 64   & 25   & 50 \\
       &  25.05.11 &	  & 6    & 87   & 54 \\
       &  14.06.11 &	  & 61   & 29   & 27 \\
167971 &  30.06.13 & prim & 3    & -16  & 40 \\
       &	   & sec  & 98   & 16   & 59 \\
       &	   & ter  & -19  & -216 & 115 \\
190918 &  25.07.10 &	  & 8    & 93   & 70 \\
\hline
\end{tabular}
\end{table}

The various components and the resulting parameters are listed in
Table~\ref{fit}. These parameters are only calculated to derive upper limits on
the magnetic field strength. They should be used with care for other studies as
they do not necessarily have a physical meaning. By running the fits
several times with different initial guess values, and by visually comparing
the quality of the fit when changing the parameters, we estimate that the
uncertainty on $v\sin i$ and $V_{\rm mac}$ is of the order of 10 km~s$^{-1}$.
The uncertainty on $V_{\rm rad}$ is of the order of a few
km~s$^{-1}$. The fits of each individual component, as well as the combined fit
of all components, are shown in Fig.~\ref{LSDprof}.

\begin{figure*}[!ht]
\begin{center}
\resizebox{\hsize}{!}{\includegraphics[clip,angle=-90]{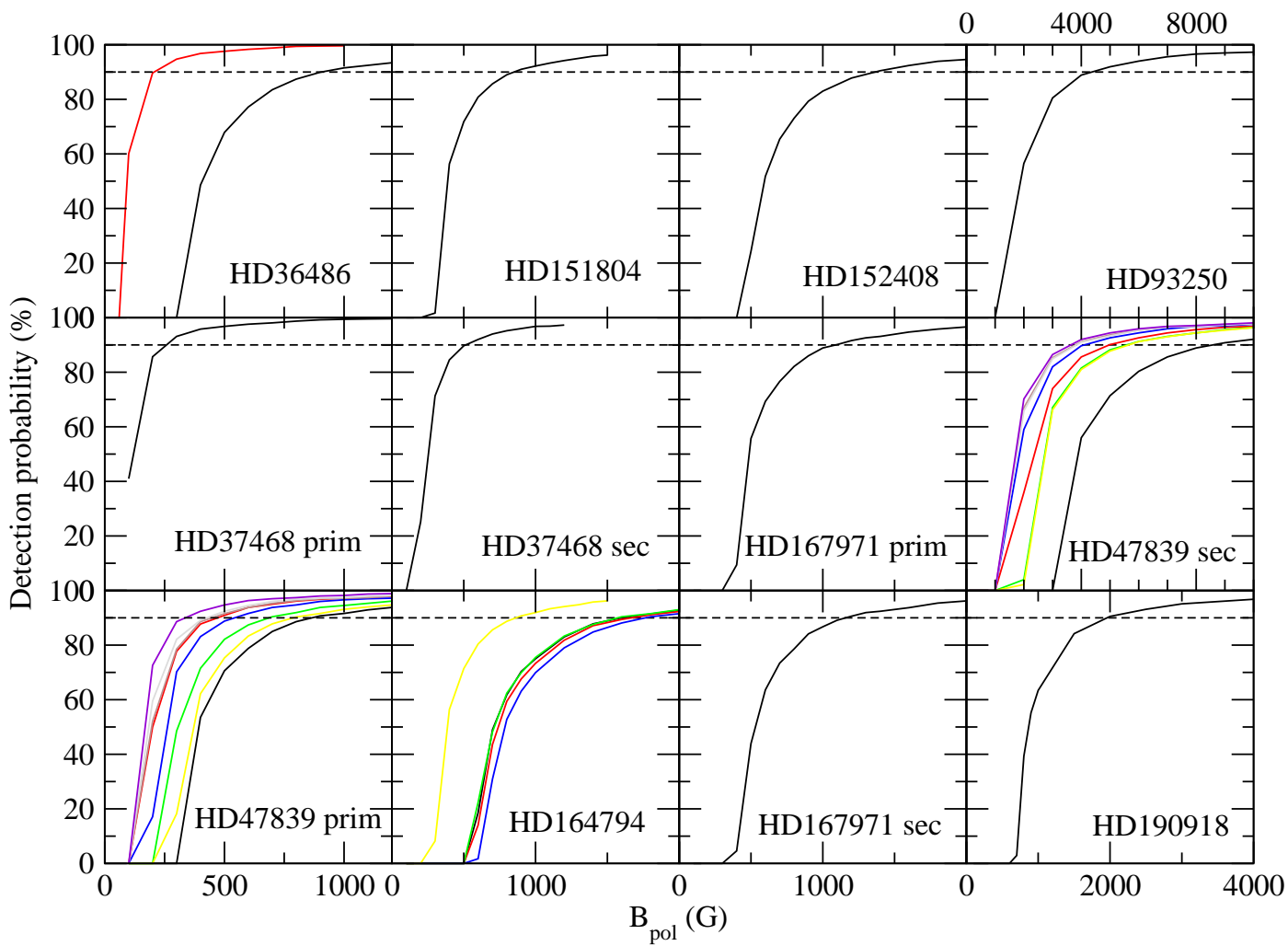}}
\caption[]{Detection probability for each spectrum of each star as a function of
the magnetic polar field strength. The horizontal dashed line indicates the 90\%
detection probability.}
\label{limit}
\end{center}
\end{figure*}

\section{Magnetic field measurements}

\subsection{Longitudinal field measurement}

From the LSD profiles we computed the longitudinal magnetic field ($B_l$) value
and the corresponding null measurement $N_l$ and their error bars $\sigma$,
using the first-order moment method of \cite{rees1979} using the form given in
\cite{wade2000}. We applied this measurement to the individual components of
each stars, when visible.

The results are reported in Table~\ref{bl}. We find that the $B_l$ and $N_l$
values are all compatible with 0 within 3$\sigma$. This confirms that no field
is detected in any of the 9 PACWB.

The magnetic field of one of our targets, HD\,93250, has already been analysed
with low-resolution FORS data by \cite{naze2012}. They did not detect a magnetic
field in this star neither, obtaining an even more stringent error bar of $\sigma\sim$80 G.

\subsection{Upper limit on undetected fields}

Since we did not detect a magnetic field signature in the 9 PACWB we studied, we
proceeded to determine the upper limit of the strength of a magnetic field that
could have remain hidden in the spectral noise.

To this aim, for various values of the polar magnetic field $B_{\rm  pol}$, we
calculated 1000 oblique dipole models of each of the LSD Stokes $V$ profiles
with random inclination angle $i$ and obliquity angle $\beta$, random rotational
phase, and a white Gaussian noise with a null average and a variance
corresponding to the SNR of each observed profile. Using the fitted LSD $I$
profiles, we calculated local Stokes $V$ profiles assuming the weak-field case
and integrated over the visible hemisphere of the star. We obtained synthetic 
Stokes $V$ profiles, which we normalised to the intensity continuum. We used the
same mean Land\'e factor (1.2) and wavelength (5000 \AA) as in the observations.

We then computed the probability of detection of a field in this set of models
by applying the Neyman-Pearson likelihood ratio test \citep[see
e.g.][]{helstrom1995,kay1998,levy2008} to decide between two hypotheses, $H_0$
and $H_1$, where $H_0$ corresponds to noise only, and $H_1$ to a noisy simulated
Stokes $V$ signal. This rule selects the hypothesis that maximises the
probability of detection while ensuring that the probability of false alarm
$P_{\rm FA}$ is not higher than a prescribed value considered acceptable.
Following values usually assumed in the literature on magnetic field detections
\citep[e.g.][]{donati1997}, we used $P_{\rm FA} = 10^{-3}$ for a marginal
magnetic detection. We then calculated the rate of detections among the 1000
models for each of the profiles of the primary and secondary stars depending on
the field strength (see Fig.~\ref{limit}).

We required a 90\% detection rate to consider that the field should have
statistically been detected. This translates into an upper limit for the
possible undetected dipolar field strength for each star and spectrum. These
upper limits are listed in Table~\ref{fit}. Since the computation of
the upper limits rely on fitted $I$ profiles, the uncertainty in the fits may
introduce an error in the field strength we derive. Comparing limits derived
from various fits of the same profile, we estimated that the error on the upper
limits could be up to $\sim$20\%.

For the 3 PACWB for which each binary component has been fitted (HD\,37468,
HD\,47839 and HD\,167971), we provide an upper limit for each star. For the
other 6 PACWB however, the result is contaminated by the undetected companion.
For two of these PACWB, either the companion has never been detected
(HD\,151804) or it is known to be a faint cool star \citep[HD\,152804,
see][]{mason1998}, and therefore the contamination can be neglected. In the case
of HD\,190918, the companion is a Wolf-Rayet star which contributes to the
spectrum with emission lines and continuum flux. Since the extracted LSD profile
is normalized to the total continuum flux, it can be treated as a single star. 

For HD\,36486, HD\,93250 and HD\,164794 however, the contribution from the
companion to the spectrum cannot be neglected. For HD\,36486 and  HD\,93250,
each component contributes to about 50\% of the flux and the $v \sin i$ values
of the primary and secondary are similar (see \cite{harvin2002} for HD\,36486
and \cite{sana2011} for HD\,93250). For these two stars, the upper limit values
should thus be considered with care and are probably underestimated by a factor
$\sim$2. For HD\,164794, the $v \sin i$ values of the two components are not
very different neither (87 and 57 km~s$^{-1}$ according to \cite{rauw2012}), but
the secondary has deeper lines than the primary. For this star too, the upper
limit value should thus be considered with care and might be significantly
underestimated.

In addition, for stars for which several observations are available, statistics
can be combined to extract a stricter upper limit taking into account that the
field has not been detected in any of the observation, using the following
equation:

\begin{equation*}
P_{\rm comb} = 100 \left[ 1 - \prod_{i=1}^n \frac{\left(100-P_i\right)}{100} \right],
\end{equation*}
where $P_i$ is the detection probability for the i$^{th}$ observation, and
$P_{\rm comb}$ is the detection probability for n observations combined. All probabilities are expressed in percents.

As an example, if two observations of one star were obtained with a detection probability of 80\% and 90\% respectively that no field stronger than 1000 G was detected, then the combined probability that such a 1000 G field was detected in none of the two observations would be 98\%.

The final upper limit derived from this combined probability for each star
for a 90\% detection probability is listed in Table~\ref{upper}.

\begin{table}
\caption{Upper dipolar field strength limit in G, combining all available data for each detected component of each star.}
\label{upper}
\begin{tabular}{lll}
\hline
\hline
Star   & Component & $B_{\rm pol,max}$ \\
       &       & G \\
\hline
HD\,36486  &      &  203  \\
HD\,37468  & prim &  258  \\
           & sec  &  513  \\
HD\,47839  & prim &  178  \\
           & sec  & 1610  \\
HD\,93250  &      & 4367  \\
HD\,151804 &      &  850  \\
HD\,152408 &      & 1363  \\
HD\,164794 &      &  605  \\
HD\,167971 & prim & 1092  \\
           & sec  & 1160  \\
           & ter  &   -   \\
HD\,190918 &      & 1960  \\
\hline
\end{tabular}
\end{table}

Finally, for one of our targets, HD\,190918, using the same ESPaDOnS spectrum as
in the present study, \cite{delachevrotiere2014} checked for the presence of a
magnetic field in the stellar wind from its emission lines. They detected
no field and determined an upper limit on the wind magnetic field of 329 G for a
95.4\% credible region using a Bayesian analysis. Their method assumes prior
knowledge on the properties of the star, in particular a pole-on orientation for
the magnetic geometry, and therefore leads to much more optimistic upper limits
than the  method presented here. The upper limit on the wind magnetic field they
obtained can therefore not be directly compared to the upper limit on the
stellar magnetic field we obtained here. 

\section{Discussion and conclusions}

\cite{parkin2014} showed that the surface magnetic field of the PACWB Cyg OB \#9
would be between 0.3 and 52 G if one assumes simple magnetic field radial
dependence, no or slow rotation, and a ratio of the energy density in the
magnetic field to the local thermal energy density ($\zeta_B$) of
$5\times10^{-5}$, or between 30 and 5200 G if that ratio is assumed to be 0.5.
In their work, the magnetic field strength scales with $\zeta_B^{1/2}$ and
$V_{\rm rot}$.

The assumptions on the field configuration and slow rotation used by
\cite{parkin2014} are probably generally not adapted to PACWB. In particular, if
the field is strong, the impact of the magnetic field on the wind, e.g.
magnetic wind confinement, should be taken into account
\citep{uddoula2002,uddoula2008}, and massive stars are often rapid rotators
\citep[e.g.][]{grunhut2013}. Nevertheless, their work provides an idea of the
typical field strengths that one might expect in PACWB.

Our analysis of archival spectropolarimetric data shows no magnetic detection in
any of the 9 PACWB for which data are available. However, the precision reached
by these archival observations is between 7 and 211 G for the measured
longitudinal field. These values are typical of the precision reached for the
measurements of fields in massive stars by the MiMeS collaboration (Grunhut et
al., in prep.). Assuming an oblique dipole field, as observed in the vast
majority of single massive stars, this leads to an upper limit of the undetected
magnetic field at 3$\sigma$ and a 90\% probability of detection between 178 and
4367 G at the stellar pole, depending on the star. 

While for some stars these archival observations are not really constraining
(e.g. HD\,93250), for several cases we can clearly exclude fields above 1000 G
and thus large $\zeta_B$ values are certainly not common in PACWB. The results
obtained for HD\,36486, HD\,37468 and HD\,47839 show that even 
dipolar fields above a few hundreds G, i.e. more moderate $\zeta_B$, do not
seem common in PACWB, while this corresponds to the typical field strength
observed in magnetic massive stars \citep{petit2013}. While the
proportion of magnetic stars among OB stars ($\sim$7\%) could fit with the
proportion of PACWB among massive binary stars, our results clearly show that
PACWB are not particularly magnetic  compared to other massive stars. 
Therefore, no link could be established between the presence of a magnetic field
typical of a magnetic massive star and the presence of synchrotron emission. 

These archival data can however not exclude fields of a few tens of G or lower.
Such field values would point towards low $\zeta_B$ values and would be
sufficient to produce synchrotron emission. However, studies of
magnetism in OB stars show that magnetic fields detected in these stars are
always relatively strong (with $B_l$ > 100 G). Weak magnetic fields are
generally not found in massive stars, even when low detection thresholds are used. This is known as the magnetic dichotomy in massive stars
\citep{auriere2007, lignieres2014}.

However, ultra weak magnetic fields have recently been detected in some A stars
\citep{lignieres2009,petit2011,blazere2014}. These fields could possibly also
exist in higher mass stars, although attempts to detect them in B stars have
been unsuccessful so far \citep{neiner2014,wadefolsom2014}. Magnetic field
amplification could exist in PACWB \citep{lucek2000, bell2001,falceta2012} and
ultra weak stellar surface magnetic field could then be sufficient to produce
synchrotron emission.

As a consequence, while this work represents the first ever effort to detect
magnetic field signatures in PACWB, provide quantitative estimates of its
possible value and constraints for models, and clearly excludes the
presence of magnetic fields typical of massive stars as the origin of
synchrotron emission in  PACWB, more precise spectropolarimetric measurements
of magnetic fields in PACWB are necessary before one can exclude the presence of
very weak magnetic fields at the surface of PACWB stars. We plan to
acquire such precise observations for very bright PACWB in the near future.

Nevertheless, even if ultra weak magnetic fields were present at the
surface of PACWB and magnetic field amplification was at work, the question
remains: if PACWB are not different, as far as their magnetic field is
concerned, from typical massive stars, why are they particle accelerators? A
possible scenario would be the production of a magnetic field at the location
of the wind shock itself.

\begin{acknowledgements}
This research has made use of the SIMBAD database operated at CDS, Strasbourg
(France), and of NASA's Astrophysics Data System (ADS). We thank the referee, M. Leutenegger, for his constructive feedback.
\end{acknowledgements}

\bibliographystyle{aa}
\bibliography{articles}

\end{document}